\documentclass[12pt]{article}
\usepackage{epsfig}
\usepackage{amsmath}
\usepackage{subfig}

\def\beq{\begin{equation}}
\def\eeq#1{\label{#1}\end{equation}}
\def\eeqn{\end{equation}}

\def\beqa{\begin{eqnarray}}
\def\eeqa#1{\label{#1}\end{eqnarray}}
\def\eeqan{\end{eqnarray}}

\let\bar=\overbar

\def\Dslash{\not{\hbox{\kern-4pt $D$}}}
\def\dslash{\not{\hbox{\kern-2pt $\del$}}}

\def\msb{{\bar{\ssstyle M \kern -1pt S}}}

\usepackage[noblocks]{authblk}
\def\Title#1{\begin{center} {\Large {\bf #1} } \end{center}}

\begin{document}

\Title{Progress of the Charged Pion Semi-Inclusive Neutrino Charged
  Current Cross Section in NOvA}



\begin{raggedright}  

{\it Aristeidis Tsaris\\
On Behalf of the NOvA Collaboration\\
Fermi National Accelerator Laboratory\\
Batavia, IL, USA}
\bigskip\bigskip
\end{raggedright}

\bigskip\bigskip





\section{Introduction}

The NOvA experiment is a long-baseline neutrino oscillation
experiment.  It uses two detectors, a near and a far detector,
placed in a high intensity neutrino beam and separated by a distance
of 810~km.   The near and far detectors are functionally identical in
detection technology, but differ in their sizes.  The near detector has a total mass of 0.3~kton while the large far
detector has a mass of 14~ktons~\cite{NOVA}.  Both detectors are placed 14~mrad off
of the central axis of the Fermilab NuMI\footnote{NuMI: Neutrinos
  at the Main Injector} beam.  The experiment is designed to
measure electron neutrino appearance and muon neutrino disappearance
rates in a muon-neutrino beam with the goals of being able to determine the neutrino mass
hierarchy, precise measurement of $\theta_{23}$ octant and establish
if there is CP violation in the lepton sector.

\begin{figure}[tb]
\begin{center}
 \centering
\epsfig{file=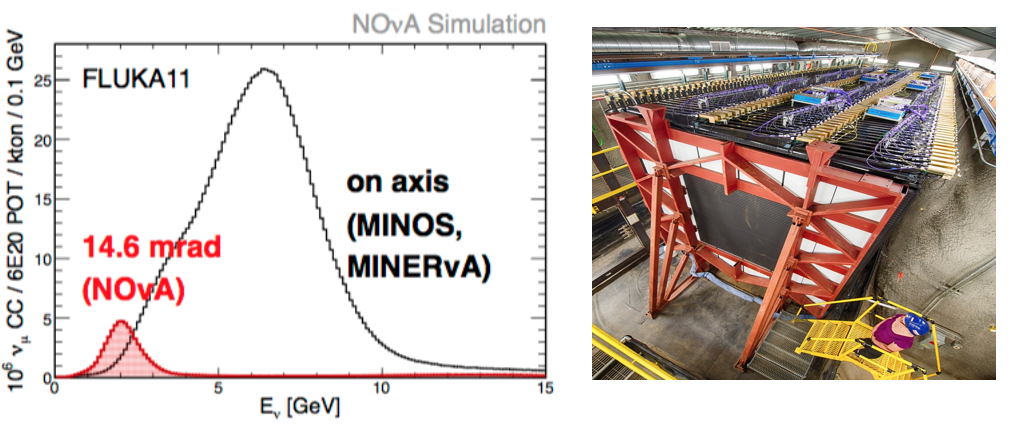,height=2.0in}
\caption{The left plot shows the neutrino energy spectrum in the NOvA
  ND (red curve) in comparison with on axis experiments (black
  curve). The right plot is a picture of the NOvA near detector.} 
\label{fig:beam}
\end{center}
\end{figure}

In addition to oscillation physics, the NOvA
experiment has a rich program of physics involving the measurement
of neutrino interaction cross sections in the near detector. The NOvA detector design has been
optimized as a low-Z tracking calorimeter, and as such is capable of
measuring both $\nu_{\mu}$ charged current interactions and $\nu_e$
charged current interactions with good energy resolution
in the near detector. The NOvA near detector (right plot of
Figure~\ref{fig:beam}) consists of 21,192
PVC plastic cells that are 3.8 m long, 3.9 cm wide and 6.6 cm
deep. They are filled with liquid scintillator which is mineral oil
with 4\% pseudocumene and comprises 62\% of the detector mass.
Each cell contains a wavelength shifting fiber (WLS) to collect the
light coming from the scintillator. Cells are arrange into plane, that alternate between
vertical (top view) and horizontal (side view) to allow for 3D
reconstruction. Light from the WLS fiber is directed onto
avalanche photo-diodes (APDs), producing an amplified electric
signal. The signal is then shaped and digitized by the Data Acquisition System.

Both NOvA detectors are exposed to neutrinos with an energy
spectrum that is centered around 2~GeV and is 14.8 milliradians off the NuMI beam axis (left plot of Figure~\ref{fig:beam}). This off-axis position gives the highest probability for oscillations, while reducing the Neutral Current background. The energy range that NOvA's near detector is exposed to is
important, as different interaction processes including quasi elastic scattering, resonant production and deep inelastic scattering contribute over different
portions of the spectrum. Performing cross section measurements
in this region yield results that can be used to tune the underlying
nuclear interaction models and can be fed back into the oscillation
measurements that NOvA is making to improve their
sensitivities. Furthermore, in this neutrino energy range the
charged current quasi elastic-scattering (CCQE) overlaps with the
resonance production and modeling the rate of energetic pion
production is important for measuring the total
reconstructed energy. Therefore, $\nu_\mu$ charged current cross
section measurements with a final state topology featuring at least
one charged pion are important for more precise oscillation
measurements.

\section{Semi-Inclusive $\nu_\mu CC(\pi^\pm)$}
In this analysis we require one muon and at least one charged pion in
the final state:
\begin{align}
    \nu_{\mu} + N \rightarrow \mu + \pi^{+/-} + X
\end{align}
Where N is the nucleus in the detector and X is the recoil nucleus
plus any other particle.

Traditionally, reconstruction of all the
particles is required prior to signal identification. In the
present work we do not require a reconstructed charged pion but rather
identify signal based on the full topology of the event. The goal of this analysis is
to measure the flux integrated double-differential cross section of
$\nu_\mu CC(\pi^\pm)$ with respect to muon kinetic energy, T, and
angle, $\theta$:
\begin{equation}
 \left(\frac{d^{2}\sigma}{d\cos\theta_{\mu}dT_{\mu}}\right)_{i} =
  \frac{\sum \mathop{}_{\mkern-5mu j} U_{ij}
    (N^{Sel}(\cos\theta_{\mu},T_{\mu})_{j} -
    N^{Bkg}(\cos\theta_{\mu},T_{\mu})_{j})
  }{\epsilon(\cos\theta_{\mu},T_{\mu})_{i} (\Delta \cos\theta_{\mu})_{i}
    (\Delta T_{\mu})_{i} N_{target}\Phi}
\end{equation}
Where $N^{Sel}$ and $N^{Bkg}$ are the number of selected and
background events respectively, $U$ is the unfolding matrix that
corrects the reconstructed distribution to the true distribution by
removing all smearing effects, $\epsilon$ is the signal selection
efficiency, $\Phi$ is the integrated neutrino flux and $N_{target}$ is
the number of targets in the fiducial volume.

\section{Event Selection}
To find muon tracks, a cluster of hits close in space and time is
reconstructed as a muon trajectory using a Kalman Filter
algorithm. Then a k-Nearest Neighbor method based on track
reconstructed observables assigns a muon score into each
track. To reject events that came from neutrino interactions in the
rock that surrounds the near detector, the start of the best muon track candidate is
required to be within a well defined fiducial volume. Additionally,
events where the muon track is not fully contained in the detector or
the hits in the cluster are not 4 cells away from the detector edges,
are removed.

\begin{figure}[tb]
\begin{center}
\epsfig{file=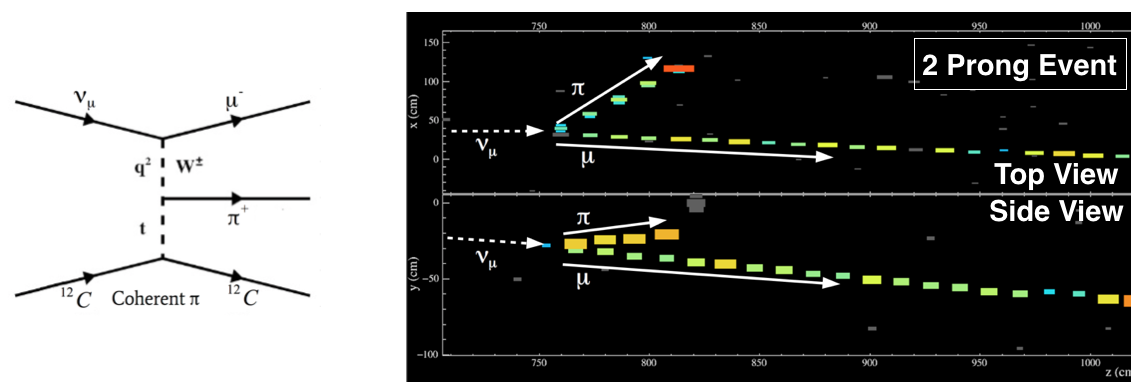,height=2.0in}
\caption{The plot on the right shows a simulated neutrino interaction
  event in the NOvA Near Detector, forming two prongs. Color represents energy deposit and
  top and side view belongs to horizontal and vertical planes
  respectively. In the left it is a schematic diagram of the coherent neutrino
  interaction that occurred.}
\label{fig:evd:2prg}
\end{center}
\end{figure}

\begin{figure}[!htb]
\begin{center}
\epsfig{file=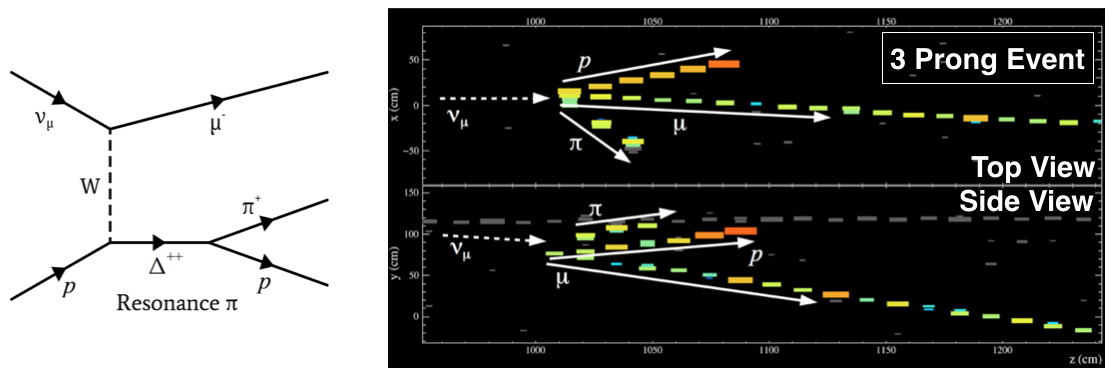,height=2.0in}
\caption{The plot on the right shows a simulated neutrino interaction
  event in the NOvA Near Detector, forming tree prongs. Color represents energy deposit and
  top and side view belongs to horizontal and vertical planes
  respectively. In the left it is a schematic diagram of the resonance neutrino
  interaction that occurred.}
\label{fig:evd:3prg}
\end{center}
\end{figure}

In NOvA "prong" is defined as a cluster of hits, close in space, following
one direction. Figure~\ref{fig:evd:2prg} and Figure~\ref{fig:evd:3prg}
show simulated events that have one charged pion in their final state
categorized by the number of prongs. The former is a
two-prong event with a pion
produced by a coherent interaction, where the neutrino interacts with
the cloud of virtual mesons surrounding the
nucleus. Figure~\ref{fig:evd:3prg} shows a typical 3 prong signal
event where the pion is coming from a $\Delta^{++}$ decay.


\section{Event Classification}
After the candidate neutrino interactions are selected, each event is
further examined to determine its likelihood of being part of the $\nu_\mu CC(\pi^\pm)$ signal sample.

To do this, we can consider final state event
identification as an image classification problem, and leverage
state of the art techniques in computing vision which have been
developed over the past decade. In particular, NOvA has
successfully developed and used deep learning techniques, denoted as
CVN~\cite{CVN}. This was used as
the primary particle identification and event classification algorithm
for electron neutrino events in the $\nu_{e}$ appearance measurements of
the 2016 datasets. CVN showed a marked improvement in the
physics sensitivity of the measurement. The increase in sensitivity
was the equivalent of approximately a 30\% increase in beam exposure.

The CVN analysis architecture that was used was inspired by the
GoogLeNet~\cite{GoogLeNet} architecture, where the two separate
detector views (XZ and YZ) of the interaction topology where mapped
into separate "color" channels, analogous to the separation RGB color
channels that is perform in photographic image recognition.  The
networks were trained using Monte Carlo simulated neutrino events,
comprised of both the signal and background, under a supervised
learning paradigm.

In this analysis, the CVN training was performed using near
detector simulated events with two targets, signal and
background. $\nu_\mu CC$ events with at least one charged pion where
used as a signal and everything else was considered as background. We
plan to use this event classification with the muon identification
algorithm defined previously, to select events for the semi-inclusive charged-pion double differential
cross section with respect to lepton kinematics.

\begin{figure}[tb]
\begin{center}
\epsfig{file=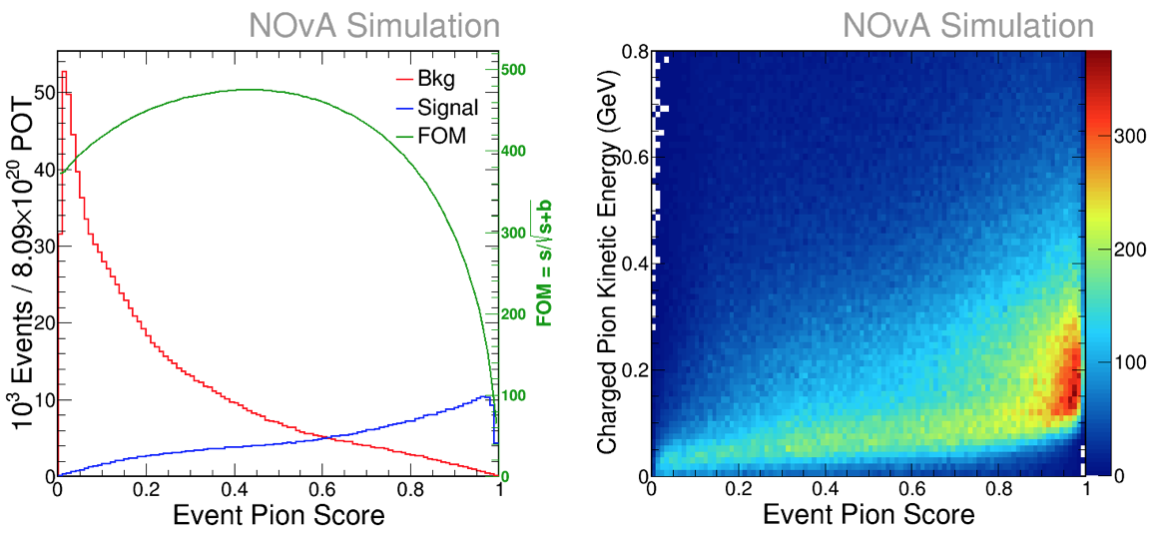,height=2.5in}
\caption{The left plot shows the score distribution of $\nu_{\mu}$ events with at least one
  charged pion in the final state. The blue curve is the signal, the
  red curve is the background and the green curve is the statistical
  figure of merit. The right plot shows the event pion score as a function of pion kinetic energy.}
\label{fig:pid}
\end{center}
\end{figure}

The left plot of Figure~\ref{fig:pid} shows the score distribution of $\nu_{\mu}$
events with at least one charged pion in the final state. We can see
a good separation of the signal versus background above 0.7. The
background curve (red curve) seems to be a 
superposition of two curves, one exponential falling for low score
values and one linear falling for higher values. The right plot of
Figure~\ref{fig:pid} shows the event score distribution as a function
of charged pion kinetic energy. As expected events with higher
energetic charged pions
form a clear band for high score values, where as for
lower energy pions we see less correlation with the event
selector. This is an indication
that the event classifier is more confident identifying events with more
energetic pions (though higher multiplicity events are expected in that region, coming
from DIS interactions).

This cross section measurement will be limited by systematic
uncertainties and so the figure of merit to select events is chosen to be the fractional
uncertainty of the cross section:
\begin{equation}
	\frac{\delta\sigma}{\sigma} = \sqrt{ \left(\frac{\delta
            N_{bkg}^{syst}}{N_{sel}-N_{bkg}}\right)^{2} + \left(\frac{\delta
            \epsilon}{\epsilon}\right)^{2} }
        \label{eq:unc}
\end{equation}
Where $\delta\epsilon$ is the fractional uncertainty in the signal efficiency
and $\delta N_{bkg}^{syst}$ is the systematic uncertainty in the
background. The systematic uncertainties that have been used for the
presented analysis are: flux uncertainties, cross section
uncertainties and energy calibration uncertainties. The
latter is a shift in the calorimetric energy of all hits by $\pm
5\%$. The event generator that was used, GENIE~\cite{Genie}, uses a
set of systematic shifts within the current experimental
uncertainties. The flux uncertainties~\cite{Leo} are mainly coming
from the poor knowledge of hadron production for the proton-nucleon
scattering in the NuMI target.

\begin{figure}[tb]
\begin{center}
\epsfig{file=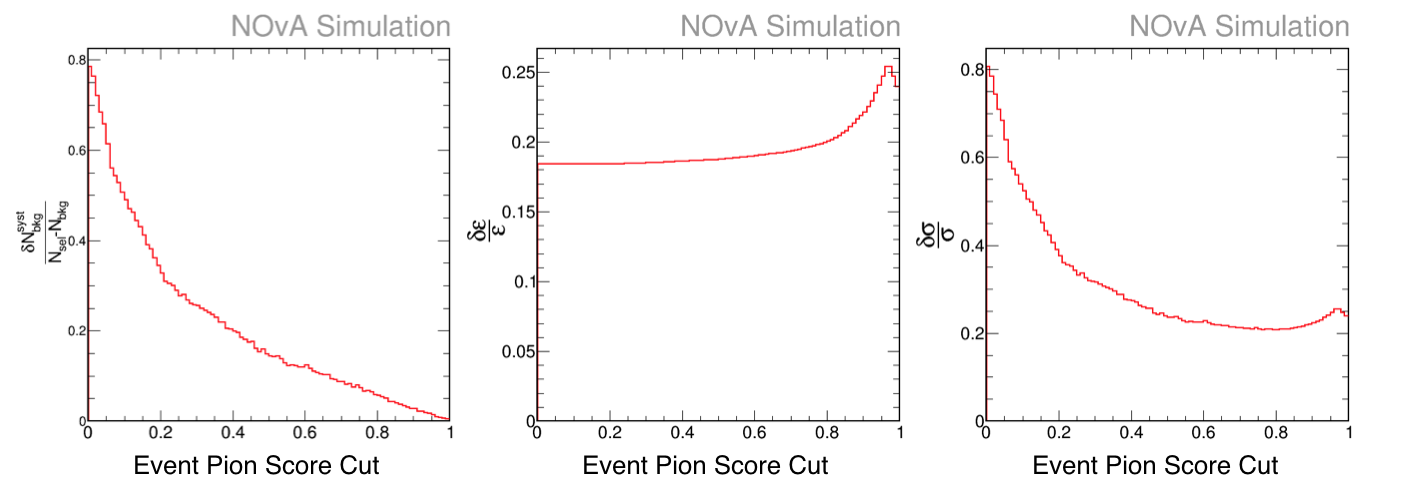,height=2.1in}
\caption{The plot on the left shows the fractional systematic uncertainty on
  background events, the middle plot shows the fractional uncertainty
  in the signal efficiency and the right plot shows the fractional
  systematic uncertainty in the cross section. The dominant systematic
  uncertainties that have been used for the presented plots are: flux
  and cross section uncertainties.}
\label{fig:unc_frac_plots}
\end{center}
\end{figure}

In order to minimize the total cross section uncertainties we want to
select events where the $\delta\sigma / \sigma$ distribution is minimized. The right and middle plot of Figure~\ref{fig:unc_frac_plots}
shows the two components of equation \ref{eq:unc}, where we see that
the fractional uncertainty in the background is falling for higher event
pion score cut values. The fractional uncertainty in the signal
efficiency appears to be flat for the most part, except in the region
of higher pion score values where it starts raising. The combination
of those two distributions can be seen in the total cross section
uncertainty, right plot of Figure~\ref{fig:unc_frac_plots},  where it
starts to fall for lower event classifier
score values and it reaches a plateau at around 0.7. Since those are
not the final systematic uncertainties that are going to be used in
this analysis without much further optimization we can temporarily use
the 0.7 as a selection value.

Table~\ref{tab:table_scores} shows the signal purity and efficiency
for events passing the classifier selection. From the selected
305,438 events, 2.9\% of those are neutral current events and 13.3\%
are $\nu_{\mu} CC 0\pi$ interaction type events. The main background
is coming
from $\nu_{\mu} CC 0\pi$ interactions and it is in line with previous
studies showing a more
difficult separation between charge pion and proton, relative to
charge pion and muon. To reduce the remaining background we are
exploring the use of kinematic variables to separate the different
particle types, requiring a michel electron in the event, and
data-driven constraints for events failing our selections. A different approach under consideration is to develop an event classifier trained on specific signal sub-categories. In that way, signal topologies that are poorly identified by the more general event classification may see a boost in selection efficiency.

\begin{table}[tb]
\begin{center}
\scalebox{0.9}
{
\begin{tabular}{l|ccccc}
 &  Cut Value &  Selected & Signal & Relative Eff(\%) & Purity(\%)  \\ \hline
 Presel &     & 2,684,460 & 740,724 & 12.4 & 27.6  \\
 CVN-pi & 0.7 &   305,438 & 237,519 & 32.1 & 77.8 \\ \hline
\end{tabular}
}
\caption{The table shows the number of events passed the selection
  criteria described in this analysis and the percentages of signal
  purity and efficiency for events with pion
  score larger than 0.7. The simulated events shown here are
  normalized to POT exposure of $8.09 \times 10^{20}$.}
\label{tab:table_scores}
\end{center}
\end{table}
\section{Conclusion}
The NOvA experiment is developing an analysis for the measurement of
charged-current $\nu_\mu$ interactions in the near detector with at
least one charged pion in the final state. This analysis, presented
here, uses a deep-learning based event classification technique. The
analysis framework optimizes the selection and 
identification criteria based on a minimization of total cross section
systematic uncertainties.

The event selection achieves a signal purity of 77.8\%, 230k signal-events
are expected to be selected, enabling a differential cross section
measurement with respect to the leading muon kinematics. The analysis
further looks to analyze events with higher energy charged pions that
can be well reconstructed by classical reconstruction techniques used
in NOvA, to perform additional measurement including the pion kinematics.

\section{Acknowledgements}
NOvA is supported by the US Department of Energy; the US National
Science Foundation; the Department of Science and Technology, India;
the European Research Council; the MSMT CR, Czech Republic; the RAS,
RMES, and RFBR, Russia; CNPq and FAPEG, Brazil; and the State and
University of Minnesota. We are grateful for the contributions of the
staffs of the University of Minnesota module assembly facility and
NOvA FD Laboratory, Argonne National Laboratory, and
Fermilab. Fermilab is operated by Fermi Research Alliance, LLC under
Contract No. DeAC02-07CH11359 with the US DOE.

\end{document}